\documentstyle[11pt,aaspp4]{article}
\slugcomment{Submitted to the Astronomical Journal}
\lefthead{Haisch et al.}
\righthead{A Mid-Infrared Study of the Young Stellar Population in the NGC 2024 Cluster}

\begin{document}

\title{A Mid-Infrared Study of the Young Stellar Population in the NGC 2024 Cluster}

\author{Karl E. Haisch Jr. \altaffilmark{1,2}, Elizabeth A. Lada
\altaffilmark{2,3}, Robert K. Pi\~{n}a \altaffilmark{2} and Charles M. Telesco \altaffilmark{2}}
\affil{Dept. of Astronomy, University of Florida, 211 SSRB, Gainesville, FL  
32611}

\and

\author{Charles J. Lada}
\affil{Smithsonian Astrophysical Observatory, 60 Garden Street, Cambridge, 
Massachusetts 02138}


\altaffiltext{1}{NASA Florida Space Grant Fellow}

\altaffiltext{2}{Visiting Astronomer at the Infrared Telescope Facility which is 
operated 
by 
the University of Hawaii under contract to the National Aeronautics and Space 
Administration.}

\altaffiltext{3}{Presidential Early Career Award for Scientists and Engineers Recipient.}


\begin{abstract}
We present the results of the first broadband {\it N} (10.8 $\mu$m) survey of the NGC 2024 cluster. The mid-infrared data were combined with our previously published {\it JHKL} photometry in order to construct spectral energy distributions (SEDs) for all detected sources. The main scientific goals were to investigate the nature of the young stellar objects (YSOs) in the cluster, and to examine the efficiency of detecting circumstellar disk sources from near-infrared {\it JHKL} color-color diagrams. Out of 59 sources surveyed having {\it K} band (2.2 $\mu$m) magnitudes m$_{K}$ $\leq$ 10.5, we detected 35 ($\sim$ 59\%) at 10 $\mu$m. Combining these detections, and upper limits for the nondetections, with existing {\it JHKL} data, we identify 1 Class I, 6 flat spectrum, 28 Class II and 5 Class III sources. We find a circumstellar disk fraction for NGC 2024 of $\sim$85\%$\pm$15\%, which confirms earlier published suggestions that the majority, if not all, of the stars in the NGC 2024 cluster formed with disks, and these disks still exist at the present time. In addition, all but one of the disk sources identified in our survey lie in the infrared excess region of the {\it JHKL} color-color diagram for the NGC 2024 cluster. This demonstrates that {\it JHKL} color-color diagrams are extremely efficient in identifying YSOs with circumstellar disks. Of the 14 sources in our survey with {\it K} -- {\it L} colors suggestive of protostellar objects, $\sim$ 29\% are protostellar in nature, while $\sim$ 7\% are true Class I sources. This may be due to extinction producing very red {\it K} -- {\it L} colors in Class II YSOs, thus making them appear similar in color to protostars. This suggests caution must be applied when estimating the sizes and lifetimes of protostellar populations within star forming regions based on {\it K} -- {\it L} colors alone. A comparison of the ratio of (Class I + flat spectrum)/(Class II + Class III) sources in NGC 2024, $\rho$ Oph, and Taurus-Auriga indicates that NGC 2024 and $\rho$ Oph have similar ages, while Taurus-Auriga is an older region of star formation, consistent with published T Tauri star ages in each region. Finally, we calculate the luminosities of the Class II sources in NGC 2024, $\rho$ Oph and Taurus and discuss the results.
\end{abstract}




%

\section{Introduction}

Understanding the early evolution of young stellar objects (YSOs) is a key ingredient for tests of our theories of the star formation process. Studies of young clusters, which likely represent the main birth places for the majority of the stars in the Galaxy (\cite{lada91}, \cite{lada92}, \cite{car00}), allow unique insights into the early phases of star formation and early stellar evolution in a statistically meaningful way. In addition, the clusters represent a different star forming environment (e.g. high stellar density, the possible presence of massive O stars) as compared to isolated star forming regions. A particularly important consideration is the potential effect of the cluster environment on the properties of the circumstellar disks which are commonly believed to surround many of the newly formed cluster members, especially since these disks represent the sites of potential planet formation.

Ideally, one would like to obtain broad band energy distributions of a significant fraction of sources in many young clusters to the hydrogen burning limit in order to study the frequency and lifetimes of the protostellar and disk phases of early stellar evolution. Unfortunately, this is impractical since observations at longer wavelengths (ie. longer than about 10 $\mu$m) are not sensitive enough to detect photospheric emission from rather massive stars. Traditionally, near-IR {\it JHK} (1.25, 1.65, 2.2 $\mu$m) color-color diagrams (i.e. a plot of {\it J-H} vs. {\it H-K}) have been used as a tool for investigating the physical natures of YSOs in young clusters (\cite{la92}; \cite{lyg93}, \cite{ll95}, \cite{lal96}). Candidate circumstellar disks are identified as objects which lie in the infrared excess region of these diagrams (\cite{la92}, \cite{mch97}). However, {\it JHK} observations alone are not long enough in wavelength to enable complete and unambiguous disk identifications. This is due, in part, to problems arising from contamination by extended emission in HII regions, reflection nebulosity, stellar photospheric emission and source crowding in high density regions. Such effects could lead to artificially high or low disk fractions as inferred from infrared excesses in {\it JHK} color-color diagrams. Furthermore, the magnitude of the near-IR excess from a disk also depends on the parameters of the star/disk system (e.g. stellar mass/age, disk inclination, accretion rate, inner disk hole size) (\cite{als87}, hereafter ALS87; \cite{mch97}; \cite{hil98}).

The magnitude of the infrared excess produced by circumstellar disks rapidly increases with increasing wavelength. Consequently, we have been conducting a program at {\it L} band (3.4 $\mu$m) to obtain a detailed and homogeneous census of circumstellar disks in a variety of young clusters with differing ages, environments and stellar contents. Recently, we have extended existing {\it JHK} observations of the NGC 2024 and Trapezium clusters to {\it L} band in order to investigate the circumstellar disk fractions in each cluster (Haisch, Lada, \& Lada (2000; hereafter HLL00), Lada et al. 2000, submitted). Very large {\it JHKL} infrared excess fractions ($\geq$80\%) were obtained to our faintest completeness limits. Assuming that the infrared excesses are produced by circumstellar disks implies that disks formed around most of the sources in these young clusters independent of stellar mass.

While it appears that the {\it JHKL} data are very efficient in identifying circumstellar disks, observations at {\it L} band are still not at a long enough wavelength to unambiguously determine the evolutionary state of the YSOs as protostellar objects, star/disk systems or sources with no circumstellar material (the Class I, II and III YSOs of ALS87). However, since contamination from photospheric emission is minimal at mid-infrared wavelengths, the presence of circumstellar disks can be ascertained from observations at 10 $\mu$m with little ambiguity if near-infrared data is also available. Therefore, combining near-infrared and mid-infrared observations represents an extremely powerful method for unambiguously identifying stars surrounded by circumstellar disks.

In this paper, we present the results of the first sensitive mid-IR 10 $\mu$m survey of the nearby young embedded cluster NGC 2024 (located in the L1630 (Orion B) giant molecular cloud at a distance of $\sim$415 pc \cite{at82}). The 59 YSOs in our magnitude limited survey (m$_{K}$ $\leq$ 10.5) were selected from our previous {\it JHK} imaging of the NGC 2024 cluster, a subset of which has been discussed in HLL00. We conducted the mid-IR imaging survey of NGC 2024 reported here in order to construct spectral energy distributions (SEDs) over a broad wavelength range to determine whether or not the excess sources identified in our previous {\it JHKL} study of the NGC 2024 cluster (HLL00) have the power law form predicted for circumstellar disks (\cite{lp74}, ALS87). Thus we investigate the efficiency of detecting circumstellar disk sources from near-IR {\it JHKL} color-color diagrams. We present the observations and reduction of our mid-IR imaging data for the NGC 2024 cluster in $\S$2. In $\S$3, we investigate the spatial distribution and physical natures of the YSOs in NGC 2024 by constructing SEDs and color-color diagrams from our near and mid-IR observations. We also calculate bolometric luminosities and extinctions for all Class II YSOs. We compare the excess fractions found from a comparison of our near and mid-IR surveys in $\S$4 and discuss the results. We summarize our primary results in $\S$5.

\section{Observations and Data Reduction}

We observed all sources in the NGC 2024 cluster with m$_{K}$ $\leq$ 10.5 at mid-infrared wavelengths. A total of 59 YSOs were included in our magnitude limited survey. Observations at {\it N} band (10.8 $\mu$m) were conducted during the periods 1996 December 13 -- 15 and 1997 September 3 -- 11 with the 3 m telescope at the NASA Infrared  Telescope Facility (IRTF) on Mauna Kea using the University of Florida 8 -- 25 $\mu$m (mid-IR) imager and spectrometer OSCIR. The array consists of a Boeing 128 $\times$ 128  pixel Si:As Blocked Impurity Band detector. The plate scale of OSCIR at the IRTF in the imaging mode is 0.223\arcsec, which gives a field of view of 28\arcsec \hspace*{0.05in}$\times$ 28\arcsec. Standard chopping and nodding techniques were used with a chop rate of 8 Hz and a 30\arcsec \hspace*{0.05in}N-S chopper throw. For all observations, the on-source integration time was 0.75 min.\footnote[4]{Further information about OSCIR is available at www.astro.ufl.edu/iag/.} 

Individual frames for each YSO were registered and combined using Interactive Data Language (IDL) routines. Many of the resulting images contained extended spatial structure due either to extended thermal emission in the source field and/or incomplete subtraction of telescope and sky background emission. This structure was modelled by masking the sources in each field and fitting a seventh order polynomial in both right ascension and declination to the unmasked portion of the image. The polynomial model was then subtracted from the image. At the time of both observing runs, thermal control of the detector array in OSCIR was achieved via closed loop heater control using a temperature sensor mounted in the cold finger assembly which attaches the detector array to the optics bench. Due to the large thermal path between the cold-finger assembly and the detector array, large variations in temperature (up to 1 K in some cases) at the array were observed using the on-array temperature sensor. This produced variations of as large as $\sim$10\% in the flat fields, although no gradients in the flat fields were obvious. (Subsequently, the closed loop temperature control was modified to use the on-array temperature sensor rather than the cold-finger temperature sensor, resulting in $\sim$10 mK temperature stability and stable flat-fields). The calibration and cluster sources were always centered in the same region of the chip, therefore errors introduced in the photometry by not flat fielding the data are estimated to be no more than a few percent. It was therefore determined that flat fielding would not improve the data and thus flat fields were not used in the reduction.

Flux calibration was performed using $\alpha$ Tau as our primary standard star and NGC 2024 IRS 2 as an internal `standard'. The in-band flux for $\alpha$ Tau was determined by integrating the spectral irradiance model from Cohen et al. (1995) through the {\it N}-band filter passband using the OSCIR filter transmission curve and an ATRAN atmospheric model for Mauna Kea. The computed in-band flux was combined with observations of $\alpha$ Tau at various airmasses to derive our flux calibration and airmass corrections. This flux was then applied to our internal 'standard' star. We observed $\alpha$ Tau at least twice on a given night. Assuming the flux of the internal `standard' remains the same at all airmasses, we calculate the calibration factors for the range of airmasses through which the cluster was observed. The fluxes were then determined for all sources detected in the cluster. The errors in the absolute flux were typically $<$10\%. Given that our observations were background limited, the 3$\sigma$ and 5$\sigma$ detection limits are 26.8 mJy and 44.7 mJy respectively.

\section{Results}

\subsection{Spatial Distribution}

We detected 35/59 (59\%) of the sources in our mid-infrared survey of the NGC 2024 cluster. Fluxes at {\it J, H, K, L}, and 10 $\mu$m are listed in Table \ref{fluxtable}. It is possible that some of the 24 nondetections are the result of limitations in the sensitivity of our survey, a point which we will examine later. The spatial distribution of all sources having {\it K} = {\it L} $\leq$ 12.0 from our {\it JHKL} survey is shown in Figure \ref{figure1} (HLL00). Three of the 59 sources were outside our {\it L} band survey region (HLL00) and are therefore not included in Figure \ref{figure1}. Superimposed on this plot is the distribution of sources from our mid-IR survey which were detected at both {\it L} band and 10 $\mu$m (star symbol) and those which were not detected at 10 $\mu$m (denoted by a pentagon). Figure \ref{figure2} shows the radial surface density profile for all mid-IR sources from our survey, both detections and nondetections, of the NGC 2024 cluster. The profile was created by counting the number of detected and nondetected sources in succesive 20'' annuli around the cluster center, taken to be the position of IRS 2, and normalizing by the annulus area. We see that, within the errors, the spatial distribution of the sources with mid-IR detections is similar to the distribution of sources not detected at 10 $\mu$m.

\subsection{Spectral Energy Distributions and Spectral Indices}

We constructed SEDs for the 35 YSOs with 10 $\mu$m detections, and each source was classified using the least squares fit to the slope between 2.2 and 10 $\mu$m. In Figure \ref{figure3}, we present the SEDs of all sources in Table \ref{fluxtable}. We calculated the spectral indices from 2.2 to 10 $\mu$m for all observed sources from the relation:

\begin{equation}
\alpha = \frac {d log(\lambda F_\lambda)}{d log \lambda}
\end{equation}

\noindent in order to quantify the natures of their SEDs (\cite{lada87}). The classification scheme of Greene et al. (1994) has been adopted in our analysis as it is believed to correspond well to the physical stages of evolution of YSOs (e.g. \cite{am94}). Class I sources have $\alpha$ $>$ 0.3, flat spectrum sources have 0.3 $>$ $\alpha$ $\geq$ --0.3, Class II sources have --0.3 $>$ $\alpha$ $\geq$ --1.6 and sources with $\alpha$ $<$ --1.6 are Class III YSOs. Table \ref{fluxtable} lists the 2.2--10 $\mu$m spectral indices, $\alpha$, for all sources and the distribution of spectral indices for the sample is shown in Figure \ref{figure4}.

An analysis of the spectral indices for the YSOs detected in NGC 2024 reveals 1 Class I source, 6 ``flat spectrum'' sources, 27 Class II sources and 1 Class III source. As noted in Greene et al. (1994), we must consider the NGC 2024 environment when classifying YSOs based on near-IR spectral indices as calculated above. The slopes of the SEDs can be steepened in regions of high extinction. Because of this, there may be more Class II and III YSOs and fewer Class I YSOs in the NGC 2024 sample than calculated using our spectral index classification scheme. We have calculated the extinction toward each object (except the Class I source) from its location in the {\it JHKL} color-color diagram (see Figure~\ref{figure7}) assuming the extinction law of Rieke \& Lebofsky (1985; hereafter RL) as discussed in section 3.4 below. Only sources 12 and 58, which become Class II and III YSOs respectively, would be reclassified. This will have a very small effect on our derived mid-infrared disk fraction (see section 3.4) since only 1 source ($\sim$ 2\% of the sample) would no longer appear to have a disk. 

\subsection{Nondetections and Sensitivity Considerations}

A total of 24/59 sources were not detected in our mid-IR survey. Three of these were not within the boundary of our {\it L} band survey. An additional source was not detected at {\it L} band and cannot be placed in the {\it JHKL} color-color diagram (see below). We therefore only consider the 20 remaining sources here. We must determine whether these sources were not detected due to sensitivity limitations or because they do not have a mid-IR excess. A Class I source with $\alpha$ = 0.3, the boundary between Class I and flat spectrum sources, would have {\it K} magnitudes of m$_{K}$ $\simeq$ 13.1 and m$_{K}$ $\simeq$ 12.6 respectively at our 3$\sigma$ and 5$\sigma$ 10 $\mu$m sensitivity limits. Therefore, since our survey included all sources brighter than a {\it K} magnitude of 10.5, we have detected all Class I sources present in our sample. We present in Figure \ref{figure5} a plot of the expected {\it N} band flux as a function of {\it K} magnitude for a Class II source with a spectral index of $\alpha$ = --1.6 (the boundary between Class II and Class III sources). At our 3$\sigma$ and 5$\sigma$ 10 $\mu$m sensitivity limits, a Class II source with $\alpha$ = --1.6 would have {\it K} magnitudes of m$_{K}$ $\simeq$ 9.9 and m$_{K}$ $\simeq$ 9.3 respectively.  These limits are labelled in Figure \ref{figure5}. In Figure \ref{figure6}, we present the distribution of sources detected at 10 $\mu$m as a function of {\it K} magnitude. The fraction of sources with mid-IR detections begins to decrease at a {\it K} magnitude of $\sim$9.0. A comparison of Figures \ref{figure5} and \ref{figure6} shows that the decrease in the fraction of 10 $\mu$m excess sources begins at approximately our 5$\sigma$ mid-IR detection limit. Therefore, for sources fainter than m$_{K}$ $\simeq$ 9.3 our nondetections may be due to sensitivity limitations rather than the YSOs not having a mid-IR excess.

Five of the 20 sources which were not detected are brighter than m$_{K}$ = 9.3. All of these sources have colors which place them in the reddening band of the {\it JHKL} color-color diagram (see Figure \ref{figure7} in next section) for the NGC 2024 cluster and are therefore classified as Class III YSOs. An examination of the {\it K} band luminosity function for our cluster and control fields reveals that these sources are likely {\it not} foreground/background stars. There are 15 sources with m$_{K}$ $>$ 9.3 which were not detected in our survey. These objects are likely either Class II or Class III, however given our sensitivity limitations, it is not possible to discriminate between the two classifications based on our mid-IR data alone. Since we are able to unambiguously detect Class I YSOs to a {\it K} magnitude of m$_{K}$ $\sim$ 13.0, the nondetections are not Class I sources. We can speculate on the nature of the 15 m$_{K}$ $>$ 9.3 nondetections by examining their near-IR colors. Eight of the fifteen sources have {\it JHKL} colors which place them in the infrared excess region of the color-color diagram (see Figure \ref{figure7} in next section). Since a near-IR excess is evidence of a YSO possibly surrounded by a circumstellar disk, these sources are candidate Class II objects. The other seven sources all lie in the reddening band and are therefore candidate Class III sources, although a Class II designation cannot be unambiguously ruled out, especially for sources which lie near the right side of the reddening band.

\subsection{Color-Color Diagrams}

In Figure \ref{figure7}, we present the {\it JHKL} color-color diagram for the sources, both detections and nondetections, in our mid-infrared survey of the NGC 2024 cluster which were within the boundaries of the HLL00 near-infrared survey. All sources are plotted showing their SED classifications. Class I sources are designated with a pentagon, flat spectrum sources with a square, Class II sources with a star and Class III sources with a triangle. The five sources determined to be Class III from our nondetections are also plotted in Figure \ref{figure7}. The remaining sources which were not detected in our survey are shown with a point. As noted in the previous section, these sources are likely either Class II or Class III, however an unambiguous determination cannot be made. In the diagram, we plot the locus of points corresponding to the unreddened main sequence as a solid line and the locus of positions of giant stars as a heavy dashed line (Bessell \& Brett 1988). The Classical T Tauri star (CTTS) locus (\cite{mch97}) is shown as a dot-dashed line. The two leftmost parallel dashed lines define the reddening band for main sequence stars and are parallel to the reddening vector. Crosses are placed along these lines at intervals corresponding to 5 magnitudes of visual extinction. The reddening law of Rieke \& Lebofsky (1985; hereafter RL) has been adopted. The adopted intrinsic colors of the latest spectral type stars observed in NGC 2024 (M5) and the location of the righthand reddening line was selected as in HLL00.

All of the Class I and flat spectrum sources lie in the infrared excess region of the {\it JHKL} color-color diagram. In addition, all but one (25/26; 96\%) of the Class II sources lie in the infrared excess region of the color-color diagram. In Figure \ref{figure8} we show the variation of {\it K} -- {\it L} with {\it K} -- {\it N}. Sources are labelled with their SED classifications as in Figure \ref{figure7}. The horizontal dashed line corresponds to the photospheric colors of an M5 main sequence star ({\it K} -- {\it L} = 0.29; \cite{bb88}). Class I sources lie to the right of the vertical dashed line, which represents sources with $\alpha$ = 0.3, while the flat spectrum and Class II sources lie to the left of this line and have {\it K} -- {\it L} $\geq$ 0.29. Class III YSOs have {\it K} -- {\it L} $\leq$ 0.29 and {\it K} -- {\it N} $\leq$ 1.0. The length of the arrow above the horizontal dashed line corresponds to the displacement produced by 10 magnitudes of visual extinction. Before being plotted in Figure \ref{figure8}, all sources except the Class I source were dereddened using the extinction law of RL. For sources in the infrared excess region of the diagram, we dereddened each source to the CTTS locus. For sources to the right of the termination point of the CTTS locus, we used adopted intrinsic colors for the Class II and flat spectrum sources. Accurate dereddened colors cannot be derived for the Class I source via this method. For the Class III sources in the reddening band of Figure \ref{figure7}, median intrinsic colors of ({\it J} -- {\it H})$_{o}$ = 0.62 and ({\it H} -- {\it K})$_{o}$ = 0.1 were adopted (\cite{ssm93}; hereafter SSM). For the Class II sources beyond the termination point of the CTTS locus, we adopted intrinsic colors of ({\it J} -- {\it H})$_{o}$ = 0.8 and ({\it H} -- {\it K})$_{o}$ = 0.5, while for flat spectrum sources ({\it H} -- {\it K})$_{o}$ = 0.75 was used (SSM; \cite{gm95}). Arrows have been placed on the 5 nondetected sources which were determined to be Class III YSOs to indicate upper limits on the {\it K} -- {\it N} colors.

There is a clear progression from the very red Class I YSO {\bf $\rightarrow$} flat spectrum {\bf $\rightarrow$} Class II {\bf $\rightarrow$} Class III. A similar trend has also been observed in $\rho$ Oph and Taurus (\cite{wly89}; \cite{kh95}). An examination of Figures \ref{figure7} and \ref{figure8} shows that source 24, the Class III source which was detected at 10 $\mu$m, has both near and mid-infrared colors indicative of a Class II YSO and may indeed be a circumstellar disk source. If this is the case, 28/40 (70\%$\pm$13\%) of the sources for which SED classes could be determined are Class II YSOs. In addition, a significant fraction of the mid-infrared flux in the flat spectrum sources (which represent a transition between Class I and Class II YSOs) can be attributed to the presence of a circumstellar disk. Therefore, including the flat spectrum sources among the circumstellar disk YSOs yields a disk fraction of 34/40 (85\%$\pm$15\%) for the NGC 2024 cluster.

\subsection{Luminosity and Extinction Estimates for Class II Sources}

In Table~\ref{lumtable}, we present bolometric luminosity and visual extinction estimates for all Class II sources in our sample. We note here that we cannot determine an accurate luminosity for the Class I and flat spectrum sources in our sample due to the paucity of our long wavelength ($\lambda$ $\leq$ 20 $\mu$m) data. We can, however, calculate bolometric luminosities for the Class II sources using the correlation between the bolometric luminosities of the Class II sources and their dereddened {\it J} and {\it K} band fluxes as derived in Greene et al. (1994). The bolometric luminosity of each source is computed by correcting the observed fluxes for interstellar extinction (A$_{V}$), which is particularly important in the NGC 2024 cluster. The A$_{V}$ values (see Table~\ref{lumtable}) were calculated using the extinction law of RL, and each source was individually dereddened to the CTTS locus. As in Greene et al. (1994), the Class II bolometric luminosities in Table~\ref{lumtable} were calculated by taking the mean of the luminosities computed from the absolute {\it J} and {\it K} band fluxes. The errors in the bolometric luminosities, $\delta$L, are equal to 1/2 the absolute value of the difference between the {\it J} and {\it K} luminosity estimates. A {\it J} band magnitude could not be determined for source 74, hence the luminosity estimate for this source is based only on the dereddened {\it K} band fluxes.

Using the empirical relationship between bolometric luminosity and {\it K} band absolute flux from Greene et al. (1994), we find that, at the distance of NGC 2024, a 1 {\it L}$_{\odot}$ Class II source would have a {\it K} band magnitude of {\it m}$_{K}$ $\simeq$ 9.9, equivalent to our 3$\sigma$ 10 $\mu$m sensitivity limit. Therefore, our observations are sensitive to the detection of a 1 {\it L}$_{\odot}$ Class II source near the front of the HII region. However, a similar Class II source embedded halfway into the HII region would have {\it m}$_{K}$ $\simeq$ 11.0 (average A$_{V}$ $\simeq$ 10.9; HLL00), and therefore would not be detected in our mid-infrared survey. We estimate that throughout most of the region surveyed, we are sensitive to the detection of Class II YSOs with {\it L} $\geq$ 3 {\it L}$_{\odot}$.

\section{Discussion}

\subsection{Comparison of Near-IR and Mid-IR Disk Fractions}

We find a circumstellar disk fraction in the NGC 2024 cluster of 85\%$\pm$15\%. This is consistent with the disk fraction of (86\%$\pm$8\%) inferred from the {\it JHKL} color-color diagram (HLL00) for all sources down to the hydrogen burning limit. In addition, all but one of the disk sources identified in our sample lie in the infrared excess region of the {\it JHKL} color-color diagram. Indeed, an almost unambiguous discrimination between disk (Class I, flat spectrum, Class II) and diskless (Class III) YSOs can be made from the {\it JHKL} colors. These results confirm that the majority, if not all, of the stars in the NGC 2024 cluster formed with disks, and these disks still exist at the present time. Our results also imply that {\it JHKL} colors are a very efficient means in which to determine circumstellar disk fractions in cluster environments.

One of our Class II sources (source 56) lies within the reddening band of the {\it JHKL} color-color diagram. It is unlikely that the colors for this source are in error since this would require larger photometric errors than those measured. It is also unlikely that variability is responsible for the location of this source in the {\it JHKL} color-color diagram since this YSO is also located in the reddening band of the {\it JHK} color-color diagram, and the {\it JHK} data were taken simultaneously (HLL00). That this object possesses a mid-IR excess suggests the presence of an inner hole in the circumstellar dust distribution. Such an object is indicative of a transition source between Class II and Class III YSOs. The lack of a near-IR excess in this source could also be due to inclination effects instead of, or in addition to, its disk having an inner hole (\cite{mch97}). Source 24, the Class III YSO ($\alpha$ = --1.9) detected at 10 $\mu$m, not only has near-IR colors which signify the presence of an infrared excess, but also has combined near and mid-IR colors typical of Class II objects (Figure \ref{figure6}). We consider this object to indeed be a Class II YSO, and suggest that the criterion for discriminating between Class II and Class III YSOs from SEDs needs to be reconsidered. In the classification scheme of Lada (1987), the boundary between Class II and III YSOs is $\alpha$ = --2.0. Given that our Class III source has $\alpha$ = --1.9 suggests that the Lada (1987) criterion may be more appropriate in making a distinction between Class II and Class III sources.

\subsection{Nature of the Very Red {\it K} -- {\it L} Sources}

Among the Taurus population of YSOs, all sources which have {\it K} -- {\it L} $\geq$ 1.5 are almost always Class I objects (\cite{kh95}). In our mid-infrared survey of NGC 2024, we find 14 sources with {\it K} -- {\it L} $\geq$ 1.5. Eight of these sources lie in the region of the {\it JHKL} color-color diagram beyond the termination of the unreddened CTTS locus, and possess extreme IR excess emission which has been attributed to the presence of candidate protostellar objects (e.g. \cite{mch97}; \cite{clada99}; HLL00; Lada et al. 2000, submitted). An additional 6 sources, located for the most part high within the CTTS reddening band, also have {\it K} -- {\it L} $\geq$ 1.5 and are thus consistent with protostellar objects.

Of the 14 sources in our survey with {\it K} -- {\it L} $\geq$ 1.5, ten sources are unambiguously identified as Class II objects and another three as flat spectrum YSOs. Therefore, of the sources with {\it K} -- {\it L} colors indicative of candidate protostellar objects, only one is indeed a true Class I source, while 3 are found to be flat spectrum protostellar sources. This suggests that $\sim$ 29\% of the {\it K} -- {\it L} candidates are protostellar in nature, while $\sim$ 7\% are true Class I sources. This may be due to the effects of extinction. It would take A$_{V}$ $\simeq$ 40 to produce a {\it K} -- {\it L} color of 1.5 magnitudes for a Class III YSO, and at least A$_{V}$ = 15 for a Class II source. Indeed 8 of the Class II sources in our sample have visual extinctions greater than A$_{V}$ = 15 (c.f. Table~\ref{lumtable}), consistent with the number of Class II YSOs which were identified as candidate protostars based on {\it K} -- {\it L} colors alone. Therefore, while an almost unambiguous identification between sources with and without disks can be made from {\it JHKL} colors, mid-infrared data are required to identify true protostellar objects. This suggests that any estimate of the numbers and lifetimes of protostellar sources based on near-IR {\it K} -- {\it L} colors should be viewed with caution.

\subsection{Comparisons with $\rho$ Oph and Taurus-Auriga}

We find 1 Class I source, 6 flat spectrum, 28 Class II and 5 Class III sources among the YSOs surveyed in the NGC 2024 cluster. Greene et al. (1994) have completed a mid-IR survey of 56 YSOs with m$_{K}$ $<$ 10 in the $\rho$ Oph cluster, supplementing previous studies of $\rho$ Oph by Wilking \& Lada (1983) and Wilking, Lada, \& Young (1989). Before we compare our results with the stellar population in $\rho$ Oph, we must account for the difference in distance between the clusters. We place the three $\rho$ Oph samples at the distance of NGC 2024 ($\sim$400 pc) and include only those sources brighter than m$_{K}$ = 10.5, as was done with our mid-IR sample. Twenty-two sources fit these criteria. 
The ratio of the number of Class I + flat spectrum sources to the number of Class II + Class III YSOs in NGC 2024 (21\% $\pm$ 8\%) is the same, within the errors, as that in $\rho$ Oph (29\% $\pm$ 13\%). This indicates that NGC 2024 and $\rho$ Oph are similar in age, consistent with the comparison of T Tauri star (TTS) ages in both NGC 2024 and $\rho$ Oph by Meyer (1996). Similarly, NGC 2024 and $\rho$ Oph are likely both younger than Taurus-Auriga since the ratio of Class I to Class II YSOs in Taurus-Auriga is smaller by at least a factor of 2 than in either NGC 2024 or $\rho$ Oph (\cite{ken90}), thus indicating that Taurus-Auriga may have a mean age which is older than both NGC 2024 and $\rho$ Oph. This is consistent with published mean ages of CTTS stars in NGC 2024 and Taurus ($\sim$ 0.3 Myr and $\sim$ 0.7 Myr) derived from the H-R diagram.

There is a progression in the mean luminosities of the Class II sources in NGC 2024, $\rho$ Oph and Taurus-Auriga. NGC 2024 has the highest mean luminosity of 7.0 $\pm$ 2.5 L$_{\odot}$, with smaller luminosities of 2.2 $\pm$ 0.3 L$_{\odot}$ and 1.3 $\pm$ 0.2 L$_{\odot}$ in $\rho$ Oph and Taurus-Auriga  respectively. We note that the mean luminosity for NGC 2024 was calculated including sources 2, 4 and 74, the three sources with luminosities much larger than the others. These three sources may be skewing the mean to anomalously high values. The median luminosity, which may be more representative of the true Class II luminosity in NGC 2024, is 2.5 $\pm$ 0.3 L$_{\odot}$. If we exclude sources 2, 4 and 74, the mean luminosity is 2.8 $\pm$ 0.3 L$_{\odot}$, similar to the median value. This is still higher than the Class II luminosity in Taurus-Auriga but similar to that in $\rho$ Oph. The mean luminosities in all three star forming regions were computed following the method of Greene et al. (1994). The mean luminosities in $\rho$ Oph and Taurus-Auriga were calculated using data from Greene et al. (1994) and Kenyon \& Hartmann (1995) respectively.

The difference in Class II luminosities between NGC 2024 and Taurus-Auriga can be understood given the difference in mean ages between the two regions. As a result of their pre-main sequence evolution, the most luminous Class II YSOs will be found in the youngest region of star formation, and the least luminous in the oldest region. The difference in the observed luminosities ($\Delta$log(L/L$_{\odot}$) $\simeq$ 0.3) does roughly correspond to that expected ($\Delta$log(L/L$_{\odot}$) $\sim$ 0.3 -- 0.4) given the difference in the mean ages of the two regions based on the pre-main sequence models of D'Antona \& Mazzitelli (1998). A higher mean luminosity in NGC 2024 relative to $\rho$ Oph can be understood if the mass distributions in the two regions are different. Indeed, a comparison of the mass functions of Class II YSOs in NGC 2024 and $\rho$ Oph by Meyer (1996) reveals that NGC 2024 is forming more massive stars. If the median luminosity in NGC 2024 is more representative, then the similarity in the Class II luminosities of NGC 2024 and $\rho$ Oph would not be surprising given that they have similar ages and accretion properties (\cite{mey96}).  

\section{Summary and Conclusions}

We have conducted the first sensitive mid-IR survey of the NGC 2024 cluster. The 59 YSOs in our magnitude limited sample (m$_{K}$ $\leq$ 10.5) were selected from our near-IR study of this star forming region (HLL00). We conducted the mid-IR imaging survey of NGC 2024 reported here in order to construct SEDs over a broad wavelength range to determine whether or not the excess sources identified in our previous published {\it JHKL} study of the NGC 2024 cluster have the power law form predicted for circumstellar disks. Thus, we not only investigate the nature of the YSOs in the cluster, but we also investigate the efficiency of detecting circumstellar disk sources from near-IR {\it JHKL} color-color diagrams. We identify 1 Class I, 6 flat spectrum, 28 Class II and 5 Class III sources based on an analysis of SEDs and combined near/mid-infrared colors. The major conclusions of our survey are summarized as follows: 

1. We find a circumstellar disk fraction in the NGC 2024 cluster of 85\%$\pm$15\%. This fraction is consistent to within the errors with the disk fraction (86\%$\pm$8\%) inferred from {\it JHKL} colors. These results confirm that the majority, if not all, of the stars in the NGC 2024 cluster formed with disks, and these disks still exist at the present time. In addition, all but one of the disk sources detected in our survey lie in the infrared excess region of the {\it JHKL} color-color diagram for NGC 2024. This suggests that {\it JHKL} color-color diagrams are a very efficient tool in estimating circumstellar disk fractions in young clusters.

2. One of our Class II sources has near-IR colors which place it in the reddening band of the {\it JHKL} color-color diagram. The location of this source in the {\it JHKL} diagram is not likely due to variability or large photometric errors. That this Class II YSO does not possess a near-IR excess suggests the presence of an inner hole in the circumstellar dust distribution, although inclination effects cannot be ruled out. 

3. The Class III source in our sample which was detected at 10 $\mu$m has near-IR colors which place it in the infrared excess region of the {\it JHKL} color-color diagram. In addition, the combined near and mid-IR colors of this source are indicative of a Class II designation rather than Class III. Therefore, this source is likely Class II and we suggest that the Lada (1987) SED index for discriminating between Class II and III sources ($\alpha$ $<$ --2.0 rather than $\alpha$ $<$ --1.6) yields better agreement with the YSO classification based on combined near/mid-infrared colors than the revised scheme of Greene et al. (1994).

4. Of the 14 candidate Class I YSOs in our mid-IR survey, only one is actually identified as a true Class I source, while 3 are found to be flat spectrum sources. This suggests that $\sim$ 29\% of the Class I candidates are protostellar in nature, while $\sim$ 7\% are true Class I YSOs. This may be the result of extinction producing very red {\it K} -- {\it L} colors in Class II YSOs, thus making them appear as candidate protostars. We therefore suggest that caution must be applied when estimating either the size or the lifetime of the protostellar population within a star forming region based only upon {\it K} -- {\it L} colors.

5. A comparison of the ratio (Class I + flat spectrum)/(Class II + Class III) sources in NGC 2024 with $\rho$ Oph and Taurus-Auriga indicates that NGC 2024 and $\rho$ Oph have similar ages, while Taurus-Auriga is an older star forming region. This is consistent with published CTTS ages in each region.

6. There is a progression in the mean luminosities of the Class II sources in NGC 2024, $\rho$ Oph and Taurus-Auriga, where the highest luminosity is found in NGC 2024, with lower luminosities in the other two regions. The difference in Class II luminosities between NGC 2024 and Taurus-Auriga can be understood given the difference in mean ages between the two regions. A higher mean luminosity in NGC 2024 relative to $\rho$ Oph can be understood if the mass distributions in the two regions are different. If the median luminosity in NGC 2024 is more representative, then the similarity in the Class II luminosities of NGC 2024 and $\rho$ Oph would not be surprising given that they have similar ages and accretion properties.

\acknowledgements
We thank the referee for providing helpful suggestions which improved the manuscript. K. E. H. gratefully acknowledges support from a NASA Florida Space Grant Fellowship and an ISO grant through JPL \#961604. E. A. L. acknowledges support from a Research Corporation Innovation Award and a Presidential Early Career Award for Scientists and Engineers (NSF AST 9733367) to the University of Florida. We also acknowledge support from WIRE grant NAG 5-6751.

\newpage

\figcaption[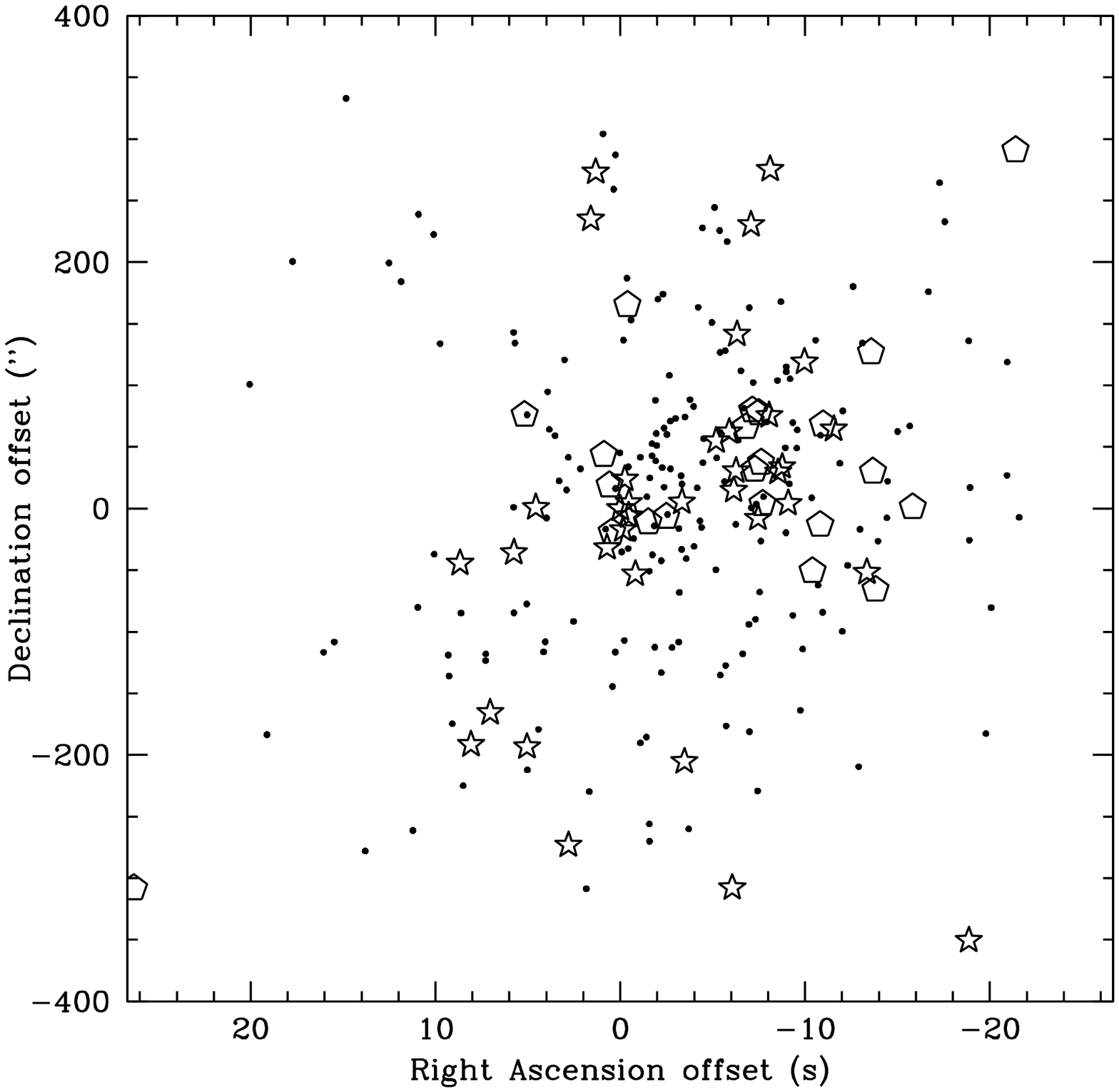]
{
Distribution of {\it L} band sources in NGC 2024. Positions of sources detected at both 3.4 and 10 $\mu$m are shown by a star while sources not detected at 10 $\mu$m are shown by a pentagon. The offsets are referred to the position $\alpha$ = 5$^{h}$41$^{m}$45$\stackrel{s}{.}$28, $\delta$ = -01$^{o}$54\arcmin31$\farcs$47 (2000).
\label{figure1}
}

\figcaption[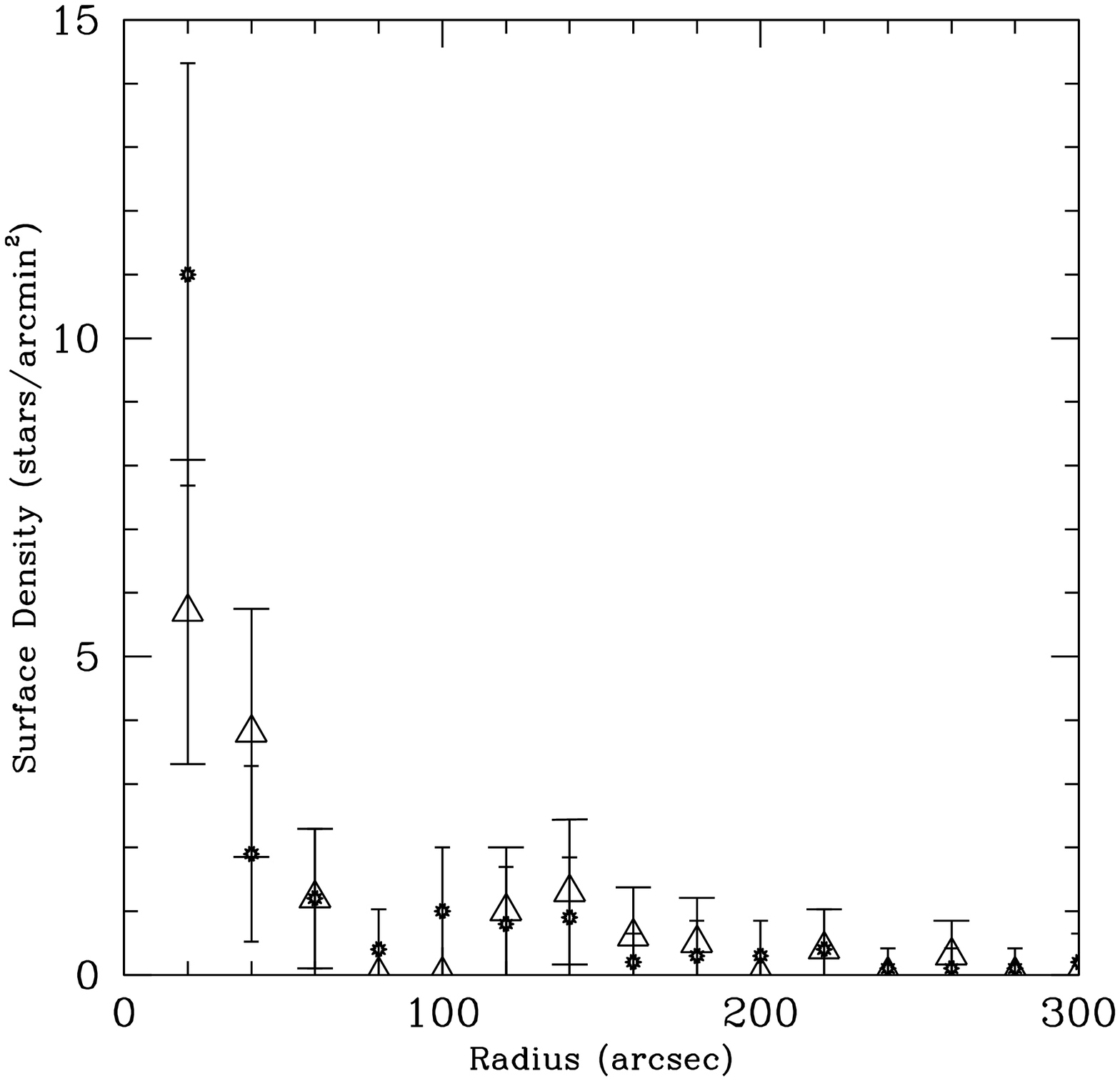]
{
Radial surface density profile for mid-IR sources, both detected and nondetected, in the NGC 2024 cluster. Sources detected at 10 $\mu$m are plotted with a filled dot and nondetected sources are plotted with a triangle. The error bars represent $\sqrt{N}$ errors in each bin.
\label{figure2}
}

\figcaption[Haisch.fig3.eps]
{
Spectral energy distributions (SEDs) of all sources detected at 10 $\mu$m. Fluxes are given in Table 1. The source ID numbers are as referenced in Table 1, and the numbers in parentheses next to each source name are the powers of 10 used to scale the SED. Figure 3a shows SEDs for $\alpha$ $>$ 0.0, Fig. 3b for 0.0 $>$ $\alpha$ $\geq$ --0.4, Fig. 3c for --0.4 $>$ $\alpha$ $\geq$ --0.9 and Fig. 3d for --0.9 $>$ $\alpha$ $\geq$ --1.9.
\label{figure3}
}

\figcaption[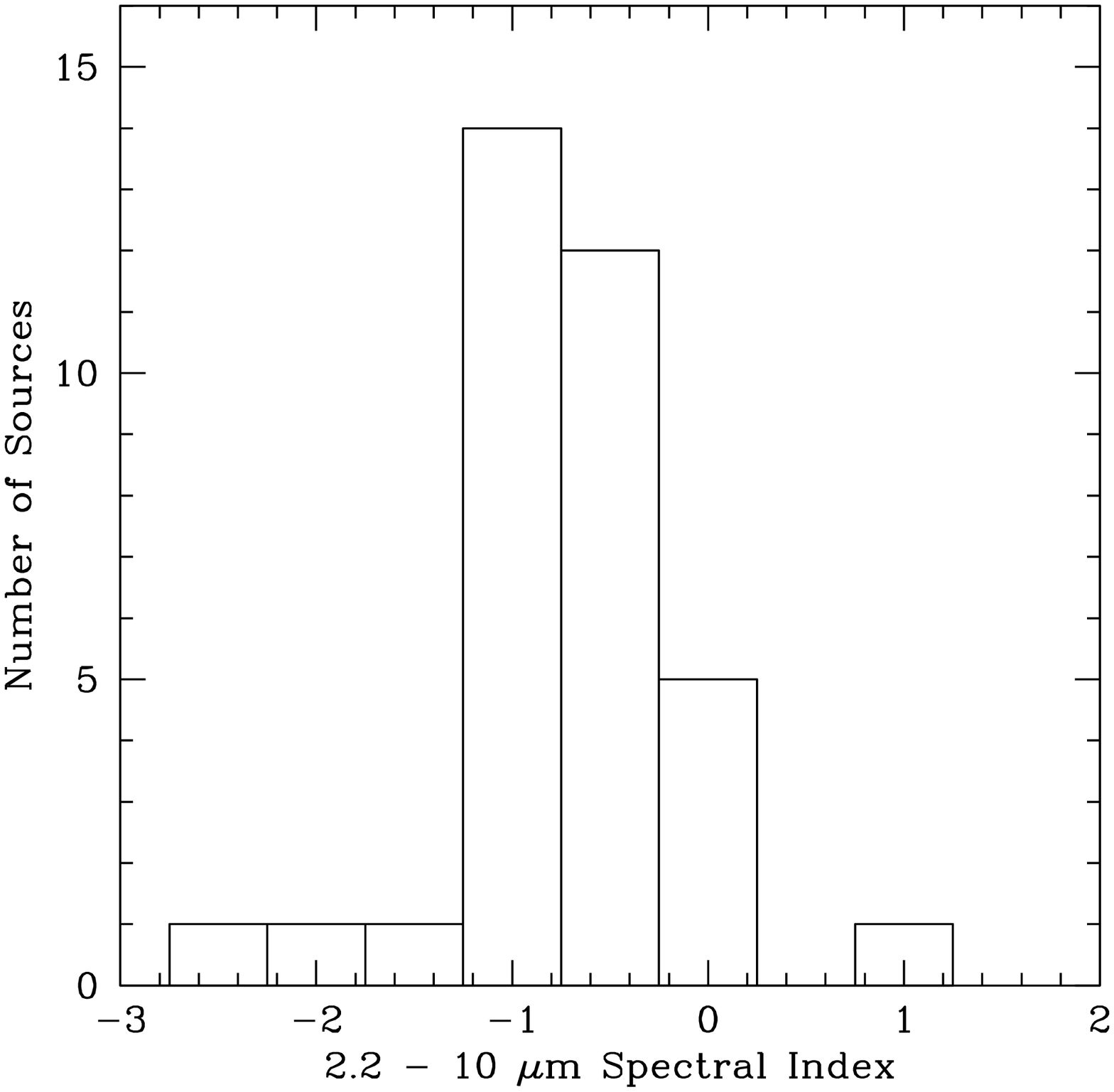]
{
The 2.2 -- 10 $\mu$m spectral index distribution for the sources detected in NGC 2024. The distribution is strongly peaked around --1.0 $<$ $\alpha$ $<$ 0.0 indicating that the majority of the detected sources are Class II objects.
\label{figure4}
}

\figcaption[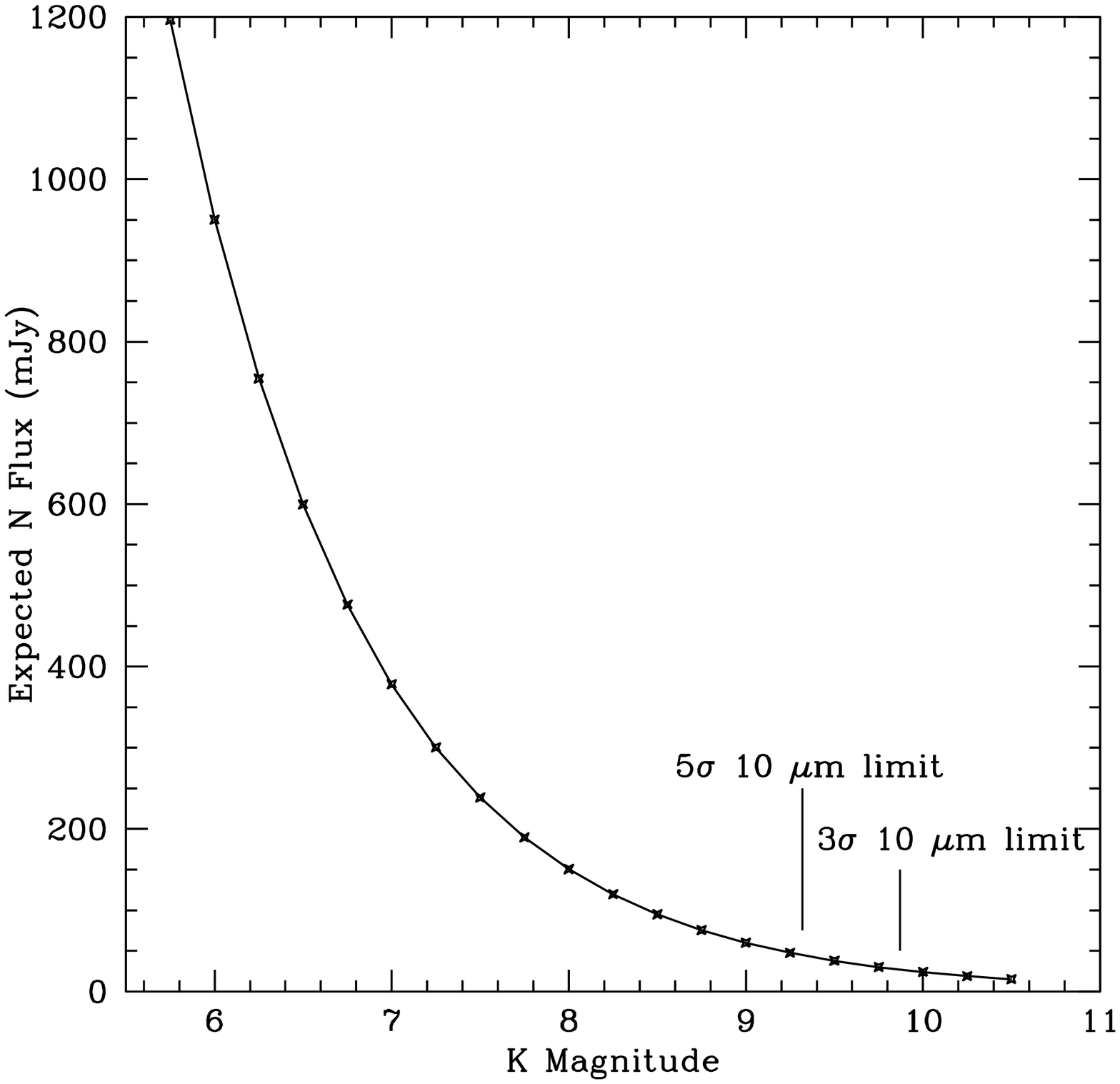]
{
Expected {\it N} band flux at a given {\it K} magnitude for Class II sources having a spectral index of $\alpha$ = --1.6. The {\it K} band magnitudes at our 3$\sigma$ and 5$\sigma$ 10 $\mu$m sensitivity limits are indicated.
\label{figure5}
}

\figcaption[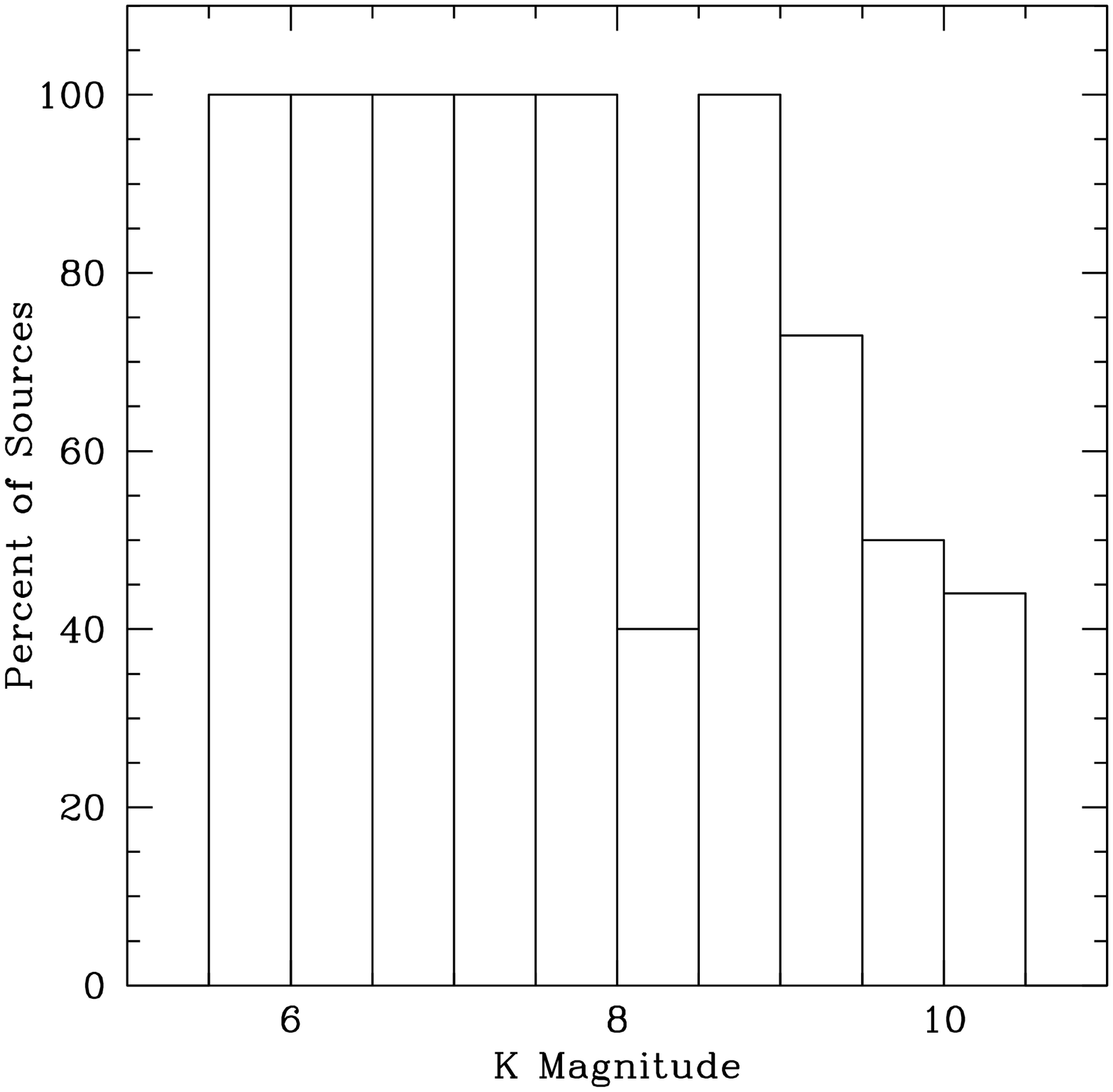]
{
Percentage of sources detected at 10 $\mu$m as a function of {\it K} magnitude. Bin size is 0.5 mag. All sources brighter than m$_{K}$=8 were detected at {\it N} band. The fraction of sources detected decreases for m$_{K}$$>$9 due to sensitivity limitations of our 10 $\mu$m data.
\label{figure6}
}

\figcaption[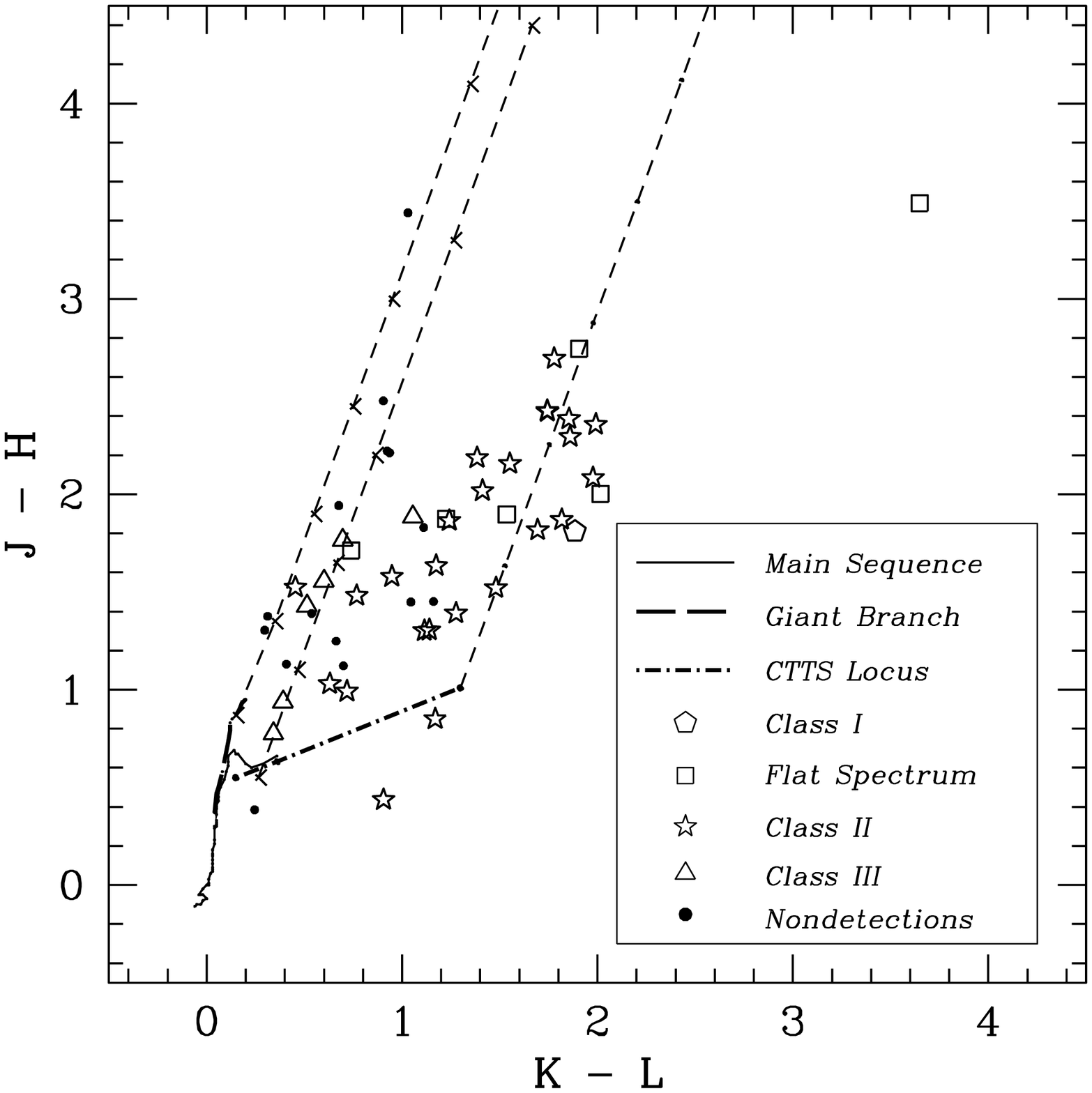]
{
{\it JHKL} color-color diagram for all sources in our 10 $\mu$m survey of NGC 2024. The sources which have {\it K} band magnitudes brighter than our 10$\mu$m sensitivity limit are plotted showing their SED classifications. Class I sources are designated with a pentagon, flat spectrum sources with a square, Class II sources with a star and Class III sources with a triangle. Sources not detected in our survey are shown with a point. In addition, we plot the locus of points corresponding to the unreddened main sequence as a solid line, the locus of positions of giant stars as a dashed line and the CTTS locus as a dot-dashed line. The two leftmost parallel dashed lines define the reddening band for main sequence stars and are parallel to the reddening vector. Crosses are placed along these lines at intervals coresponding to 5 mag of visual extinction. The rightmost dashed line is parallel to the reddening band.
\label{figure7}
}

\figcaption[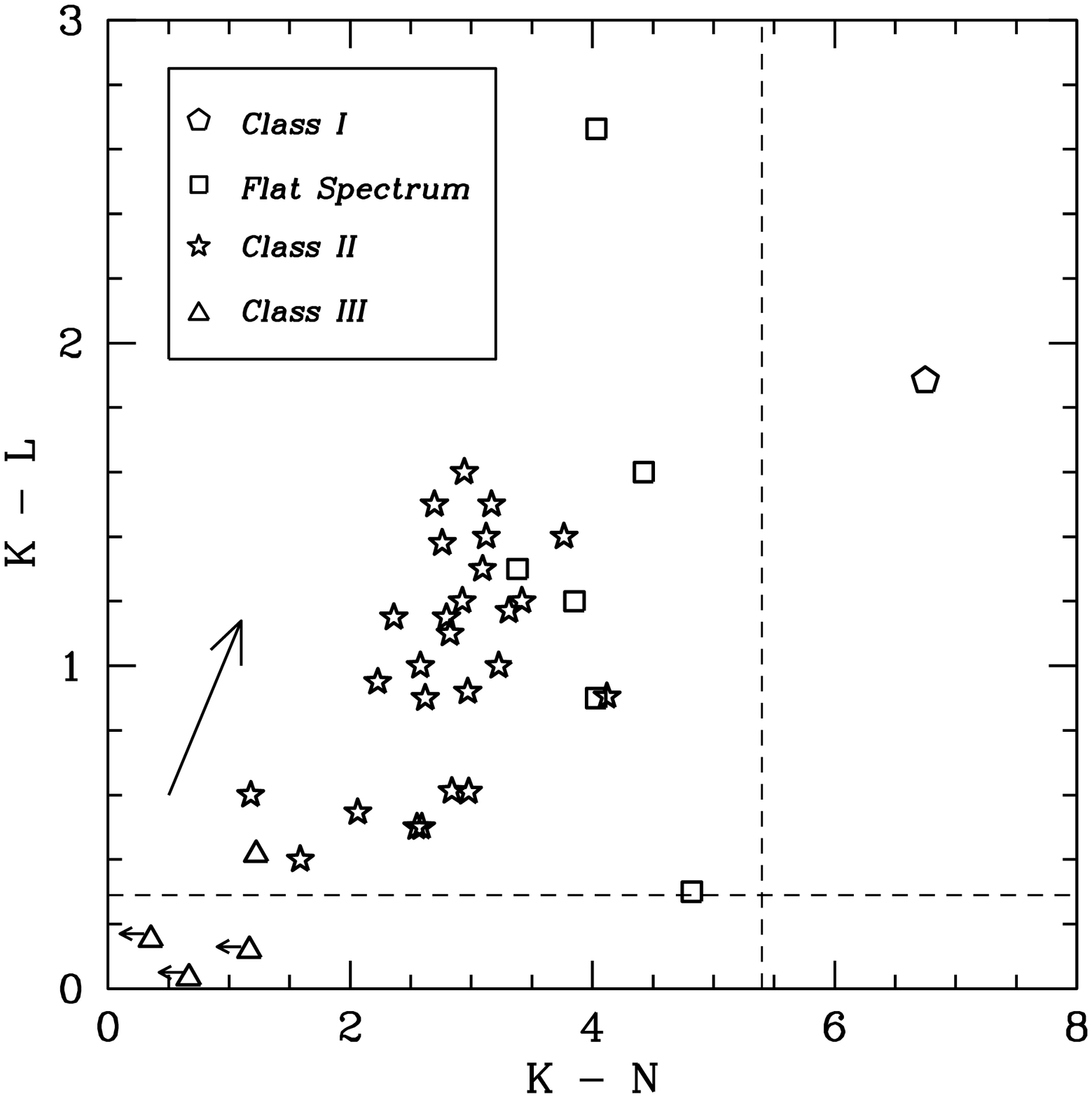]
{
Color-color diagram showing variation of {\it K} -- {\it L} with {\it K} -- {\it N} for the sources with {\it K} magnitudes brighter than our 10$\mu$m sensitivity limit. Class I sources are designated with a pentagon, flat spectrum sources with a square, Class II sources with a star and Class III sources with a triangle. All sources except the Class I source have been dereddened. The horizontal dashed line corresponds to a star with M5 main sequence colors. The vertical line represents the {\it K} -- {\it N} color for a source with $\alpha$ = 0.3. The length of the arrow above the horizontal dashed line corresponds to the displacement produced by 10 magnitudes of visual extinction. Arrows on the 5 nondetected sources which were determined to be Class III YSOs indicate upper limits on the {\it K} -- {\it N} colors. There appears to be a smooth trend from the Class I source to the bluest Class II sources. The Class III source which has near-infrared colors indicative of circumstellar disks lies within the region occupied by the Class II sources and should likely be reclassified (see text).
\label{figure8}
}

\clearpage
\begin{deluxetable}{ccccccccc}
\footnotesize
\tablecaption{{\it JHKL} and 10$\mu$m Fluxes for Sources Surveyed in NGC 2024
\label{fluxtable}}
\tablewidth{0pt}
\tablehead{Source ID & R.A.(2000) & Dec (2000) & $\alpha$ & J & H & K & L & N \nl
 & & & (2.2-10 $\mu$m) & (mJy) & (mJy) & (mJy) & (mJy) & (mJy)}
\startdata
1  & 5 41 45.79 & -01 54 31.43 & 0.2 & 60.0 & 910.0 & 3100.0 & 41900.0 & 19000.0\nl
2  & 5 41 37.79 & -01 54 39.05 & -0.4 & 1100.0 & 1050.0 & 1550.0 & 1700.0 & 3800.0\nl
4  & 5 41 39.04 & -01 52 09.74 & -0.4 & 43.1 & 220.0 & 650.0 & 1500.0 & 2100.0\nl
6  & 5 41 49.79 & -01 54 30.92 & 0.2 & 120.0 & 330.0 & 580.0 & 520.0 & 3400.0\nl
12 & 5 41 37.28 & -01 49 55.71 & -0.3 & 29.6 & 100.0 & 250.0 & 450.0 & 800.0\nl
14 & 5 41 37.22 & -01 53 15.51 & -1.2 & 130.0 & 190.0 & 250.0 & 200.0 & 190.0\nl
22 & 5 41 44.79 & -01 54 26.56 & -0.7 & 9.7 & 61.8 & 190.0 & 420.0 & 380.0\nl
23 & 5 41 46.68 & -01 49 58.79 & -1.3 & 66.5 & 130.0 & 190.0 & 240.0 & 140.0\nl
24 & 5 41 39.00 & -01 54 00.54 & -1.9 & 21.8 & 79.3 & 190.0 & 200.0 & 53.9\nl
26 & 5 41 36.19 & -01 54 26.83 & -0.7 & 19.2 & 61.8 & 170.0 & 380.0 & 320.0\nl
29 & 5 41 50.98 & -01 55 07.61 & 1.0 & 23.3 & 67.8 & 150.0 & 380.0 & 3900.0\nl
33 & 5 41 38.27 & -01 50 40.88 & -1.1 & 21.2 & 73.0 & 140.0 & 180.0 & 130.0\nl
38 & 5 41 41.73 & -01 57 56.54 & -0.2 & 16.4 & 60.7 & 130.0 & 190.0 & 480.0\nl
39 & 5 41 33.71 & -01 53 26.25 & -1.1 & 42.3 & 88.6 & 130.0 & 180.0 & 120.0\nl
40 & 5 41 39.09 & -01 59 38.85 & -1.1 & 80.7 & 120.0 & 130.0 & 110.0 & 100.0\nl
44 & 5 41 45.09 & -01 54 48.28 & -0.2 & 3.4 & 23.3 & 120.0 & 300.0 & 490.0\nl
45 & 5 41 39.20 & -01 54 16.15 & -0.9 & 3.6 & 31.9 & 120.0 & 310.0 & 180.0\nl
48 & 5 41 47.90 & -01 59 04.79 & 0.0 & 11.3 & 42.8 & 110.0 & 310.0 & 610.0\nl
49 & 5 41 39.41 & -01 53 28.86 & -1.2 & 5.2 & 39.7 & 100.0 & 160.0 & 88.1\nl
53 & 5 41 36.80 & -01 54 00.78 & -1.2 & 22.0 & 60.7 & 100.0 & 81.4 & 71.0\nl
54\tablenotemark{1} & 5 41 26.28 & -02 00 20.62 & -0.6 & 41.9 & 71.0 & 100.0 & -- & 180.0\nl
56 & 5 41 36.50 & -01 53 56.85 & -1.1 & 26.0 & 72.3 & 97.4 & 62.3 & 71.6\nl
58 & 5 41 44.38 & -01 55 24.73 & -1.6 & 9.6 & 34.0 & 95.6 & 150.0 & 46.5\nl
61 & 5 41 40.11 & -01 53 36.02 & -0.8 & 4.4 & 23.9 & 93.0 & 200.0 & 150.0\nl
64 & 5 41 31.98 & -01 55 22.03 & -0.9 & 19.2 & 49.6 & 80.2 & 83.6 & 94.4\nl
73 & 5 41 44.79 & -01 54 37.06 & -1.0 & 3.6 & 15.0 & 57.1 & 100.0 & 63.5\nl
74 & 5 41 45.00 & -01 54 07.57 & -0.9 & $<$3.2 & 4.7 & 57.1 & 190.0 & 88.3\nl
75 & 5 41 52.24 & -01 57 17.60 & -0.9 & 12.0 & 31.9 & 56.5 & 70.9 & 68.2\nl
78 & 5 41 53.23 & -01 57 43.57 & -0.4 & 7.0 & 21.8 & 54.0 & 110.0 & 170.0\nl
80 & 5 41 41.89 & -01 54 25.95 & -0.6 & 3.2 & 15.7 & 53.5 & 130.0 & 120.0\nl
88 & 5 41 53.88 & -01 55 16.82 & -0.8 & 4.2 & 17.2 & 46.6 & 120.0 & 75.7\nl
90 & 5 41 46.87 & -01 50 36.90 & -0.9 & 14.3 & 31.6 & 45.8 & 55.3 & 59.3\nl
92 & 5 41 45.98 & -01 55 03.04 & -0.6 & $<$3.2 & 9.7 & 44.5 & 97.8 & 95.0\nl
98 & 5 41 50.23 & -01 57 45.15 & -1.0 & 6.3 & 19.7 & 41.7 & 52.3 & 44.0\nl
100 & 5 41 35.33 & -01 52 31.87 & -0.9 & 7.9 & 23.1 & 40.6 & 68.9 & 54.2\nl
\enddata
\tablenotetext{1}{Source is not in our {\it L} band survey region.}
\end{deluxetable}

\clearpage
\begin{deluxetable}{cccc}
\small
\tablecaption{Luminosities and A$_{V}$ for Class II Sources\tablenotemark{1} \label{lumtable}}
\tablewidth{0pt}
\tablehead{Source ID & A$_{V}$ (mag) & L (L$_{\odot}$) & $\delta$L (L$_{\odot}$)}
\startdata
2 & 8.7 & 35.0 & 11.0\nl
4 & 14.0 & 30.0 & 0.2\nl
14 & 0.8 & 3.3 & 0.3\nl
22 & 13.0 & 7.5 & 0.9\nl
23 & 4.0 & 4.0 & 0.7\nl
24 & 8.5 & 5.0 & 0.3\nl
26 & 7.5 & 3.5 & 0.8\nl
33 & 9.4 & 4.7 & 0.8\nl
39 & 5.4 & 3.2 & 0.6\nl
40 & 1.6 & 2.5 & 0.6\nl
45 & 12.0 & 3.7 & 0.8\nl
49 & 12.0 & 4.3 & 0.2\nl
53 & 5.6 & 2.3 & 0.3\nl
54 & 4.6 & 2.8 & 0.6\nl
56 & 5.6 & 2.8 & 0.9\nl
58 & 8.8 & 2.3 & 0.6\nl
61 & 11.0 & 2.2 & 0.7\nl
64 & 7.1 & 2.5 & 0.5\nl
73 & 11.0 & 1.8 & 0.5\nl
74 & 43.0 & 61.0 & --\nl
75 & 7.1 & 1.6 & 0.1\nl
78 & 6.5 & 1.1 & 0.3\nl
80 & 12.0 & 1.7 & 0.5\nl
88 & 9.8 & 1.3 & 0.3\nl
90 & 4.9 & 1.1 & 0.1\nl
92 & 14.0 & 1.7 & 0.5\nl
98 & 6.1 & 1.0 & 0.7\nl
100 & 5.5 & 1.0 & 0.1\nl
\enddata
\tablenotetext{1}{Luminosity estimates, errors and A$_{V}$ values are calculated
as in Greene et al. (1994) with all sources dereddened to the CTTS locus.}
\end{deluxetable}

\clearpage

\clearpage
\plotone{Haisch.fig1.eps}
\clearpage
\plotone{Haisch.fig2.eps}
\clearpage
\plotone{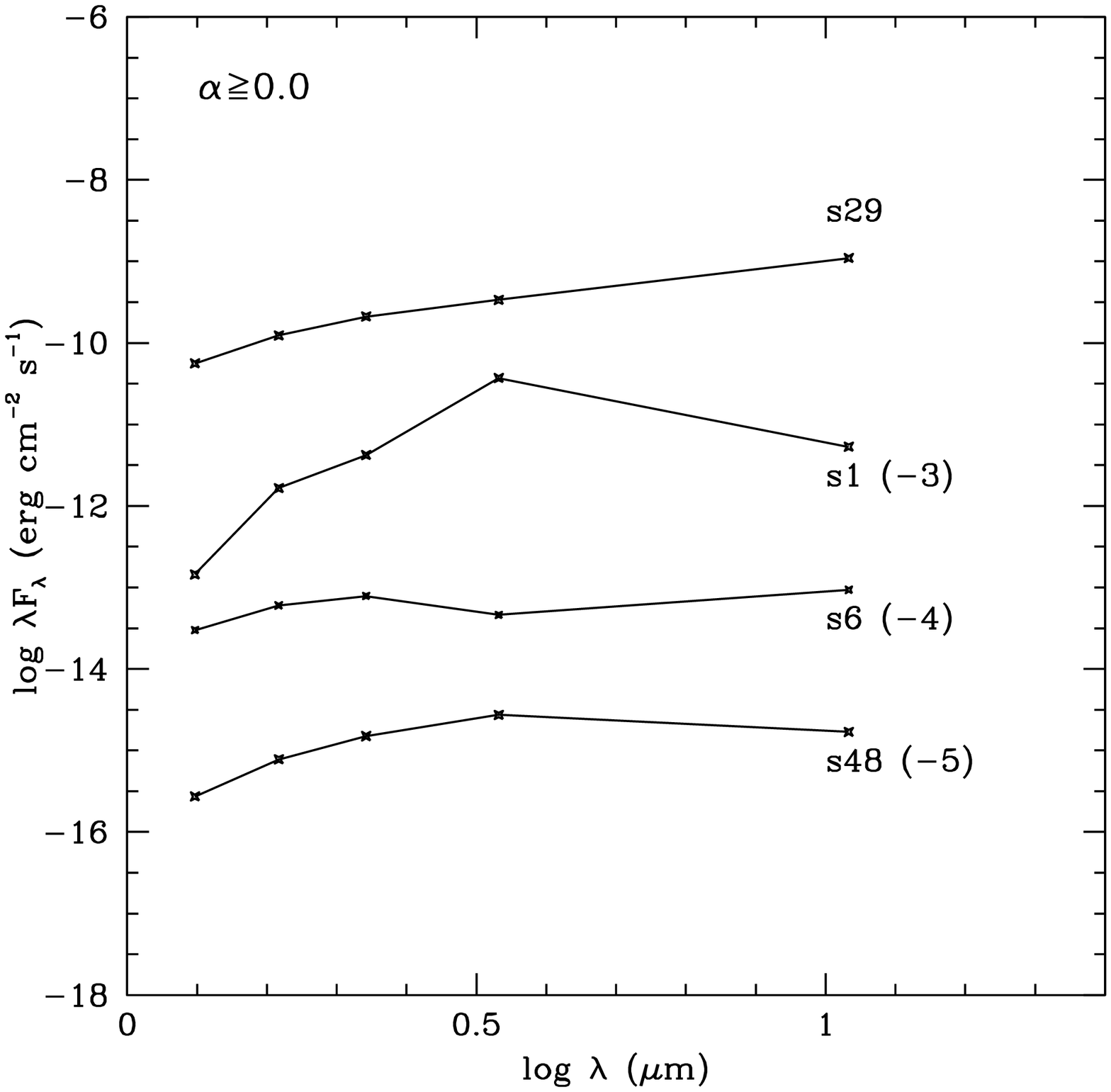}
\clearpage
\plotone{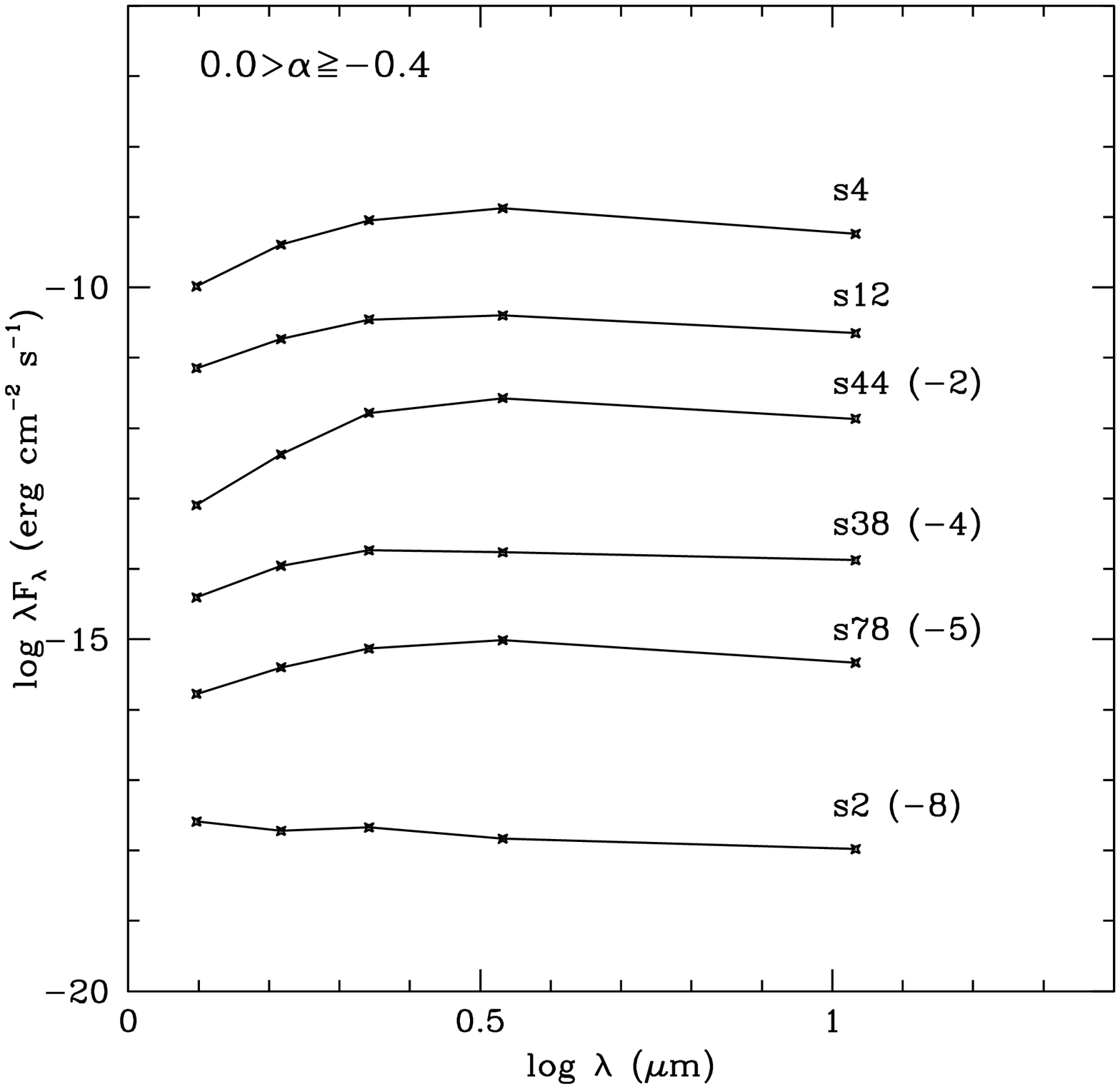}
\clearpage
\plotone{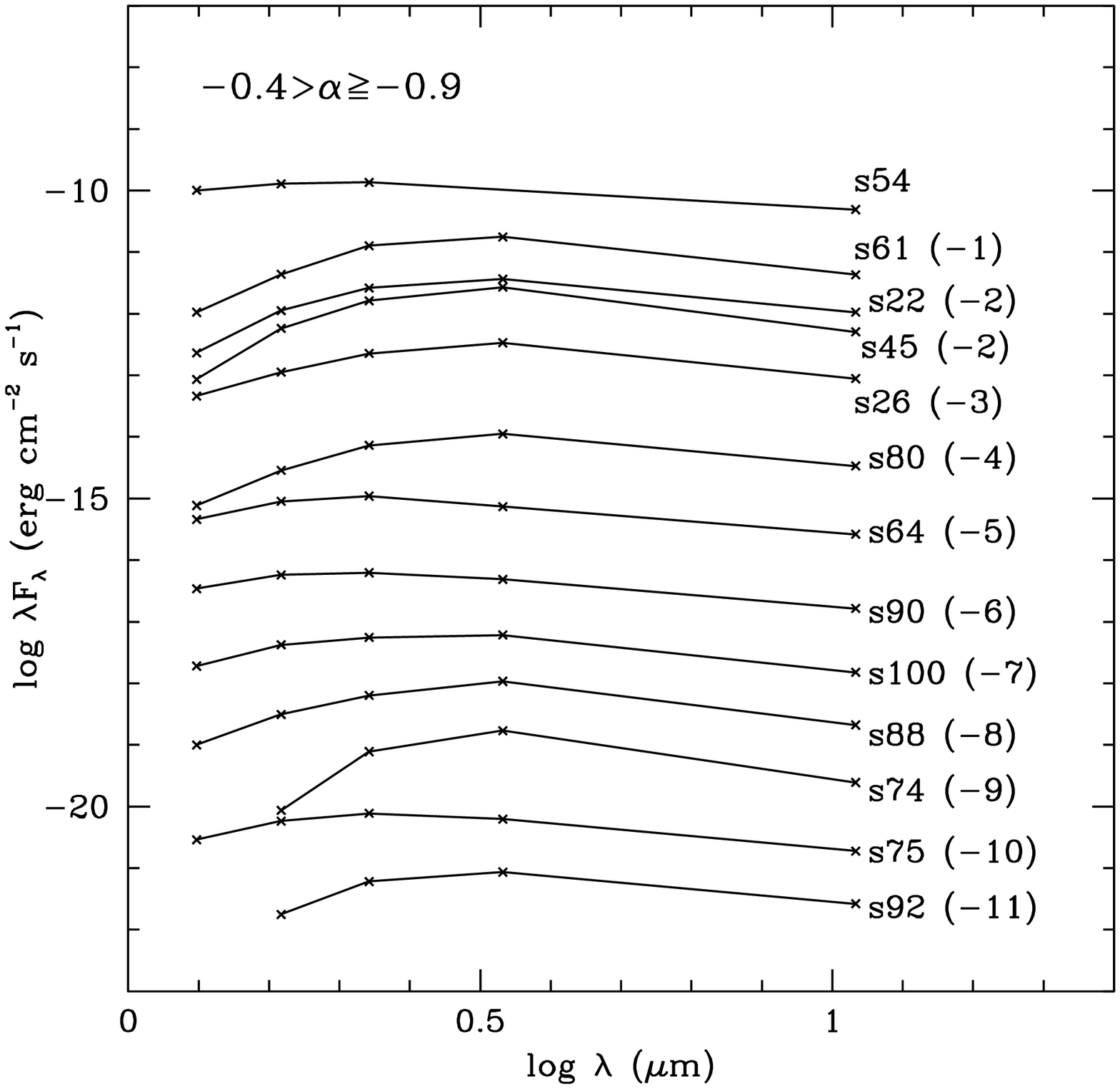}
\clearpage
\plotone{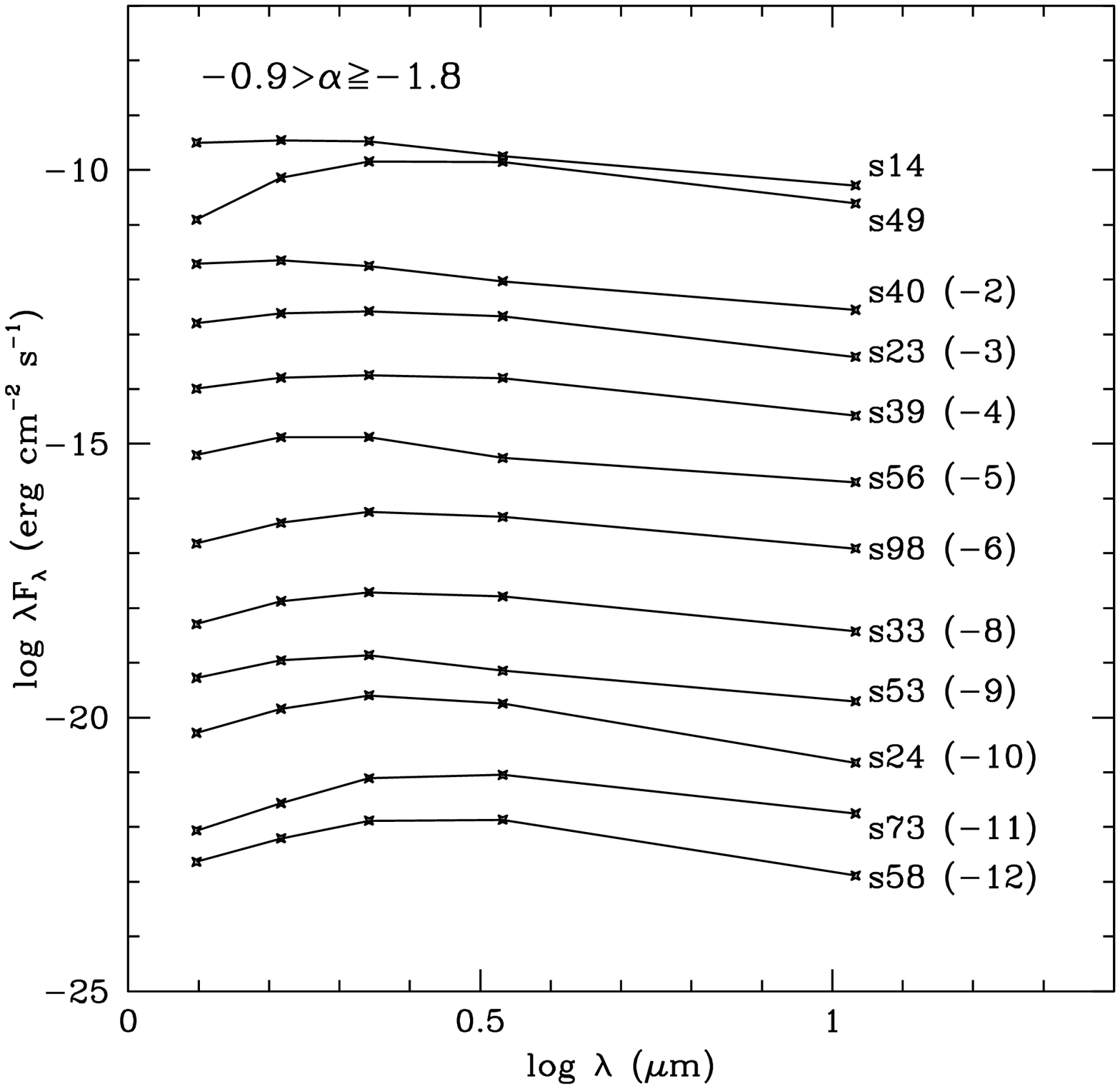}
\clearpage
\plotone{Haisch.fig4.eps}
\clearpage
\plotone{Haisch.fig5.eps}
\clearpage
\plotone{Haisch.fig6.eps}
\clearpage
\plotone{Haisch.fig7.eps}
\clearpage
\plotone{Haisch.fig8.eps}
\clearpage

\end{document}